\documentclass[conference]{IEEEtran}

\usepackage{mathptmx}
\usepackage{graphicx}
\usepackage{times}
\usepackage{caption}
\usepackage{subcaption}
\usepackage[hidelinks]{hyperref}
\usepackage{amsmath}
\usepackage{flushend}
\usepackage{array}
\usepackage[ruled]{algorithm2e}

\ifCLASSINFOpdf
\else
\fi
\renewcommand\IEEEkeywordsname{Keywords}

\begin{document}
%
\title{Towards Developing an Easy-To-Use Scripting Environment for Animating Virtual Characters}

\author{\IEEEauthorblockN{Christos Mousas}
\IEEEauthorblockA{Department of Computer Science\\
Southern Illinois University\\
Carbondale, IL 62901, USA\\
Email: christos@cs.siu.edu}
}


%


\maketitle

\begin{abstract}
This paper presents the three scripting commands and main functionalities of a novel character animation environment called CHASE. CHASE was developed for enabling inexperienced programmers, animators, artists, and students to animate in meaningful ways virtual reality characters. This is achieved by scripting simple commands within CHASE. The commands identified, which are associated with simple parameters, are responsible for generating a number of predefined motions and actions of a character. Hence, the virtual character is able to animate within a virtual environment and to interact with tasks located within it. An additional functionality of CHASE is supplied. It provides the ability to generate multiple tasks of a character, such as providing the user the ability to generate scenario-related animated sequences. However, since multiple characters may require simultaneous animation, the ability to script actions of different characters at the same time is also provided.
\end{abstract}


%
\IEEEpeerreviewmaketitle

\section{Introduction}
\label{sec1}
Character animation can be characterized as a complex and time-consuming process. This is especially true when animating virtual characters based on key-frame techniques, as this requires prior knowledge of software solutions. Moreover, artistic skills are also required since the virtual character should animate as naturally as possible. 

In order to avoid time-consuming processes in animating virtual characters, motion capture technologies now provide high quality and realistic animated sequences. This is possible because the ability to capture real humans in the act of performing is achieved through the provided required motions. The advantages of motion capture techniques are numerous, especially in the entertainment industry. However, the captured motion data, itself, it is not always usable, since virtual characters should be able to perform tasks in which the required constraints are not always fulfilled. Thus, methodologies that retarget \cite{ref1}, wrap \cite{ref2}, blend \cite{ref3}\cite{ref4}, splice \cite{ref5}\cite{ref43}\cite{ref44}, interpolate \cite{ref6}\cite{ref7} etc., the motion data have become available to help the animators to create the required motion sequences. In addition to the motion synthesis techniques that are based on software solutions, animating a virtual character through programming is also difficult. This is especially true in cases where animators, artists and students do not have the required programming skills. Hence, animating virtual characters in order to visualize ideas and generate simple scenarios in which virtual characters evolve can be a very complex process.

Based on the aforementioned difficulties that inexperienced programmers can face, this paper introduces a simple, easy-to-use, scripting environment for animating virtual characters, which is based on a small number of scripting commands. The scripting environment presented (see Fig. 1), which is called CHASE, provides a user with the ability to script the action of a character as well as to script possible interaction between a character and objects that are located within the virtual environment.

\begin{figure*}[!t]
\centering
\includegraphics[width=\textwidth]{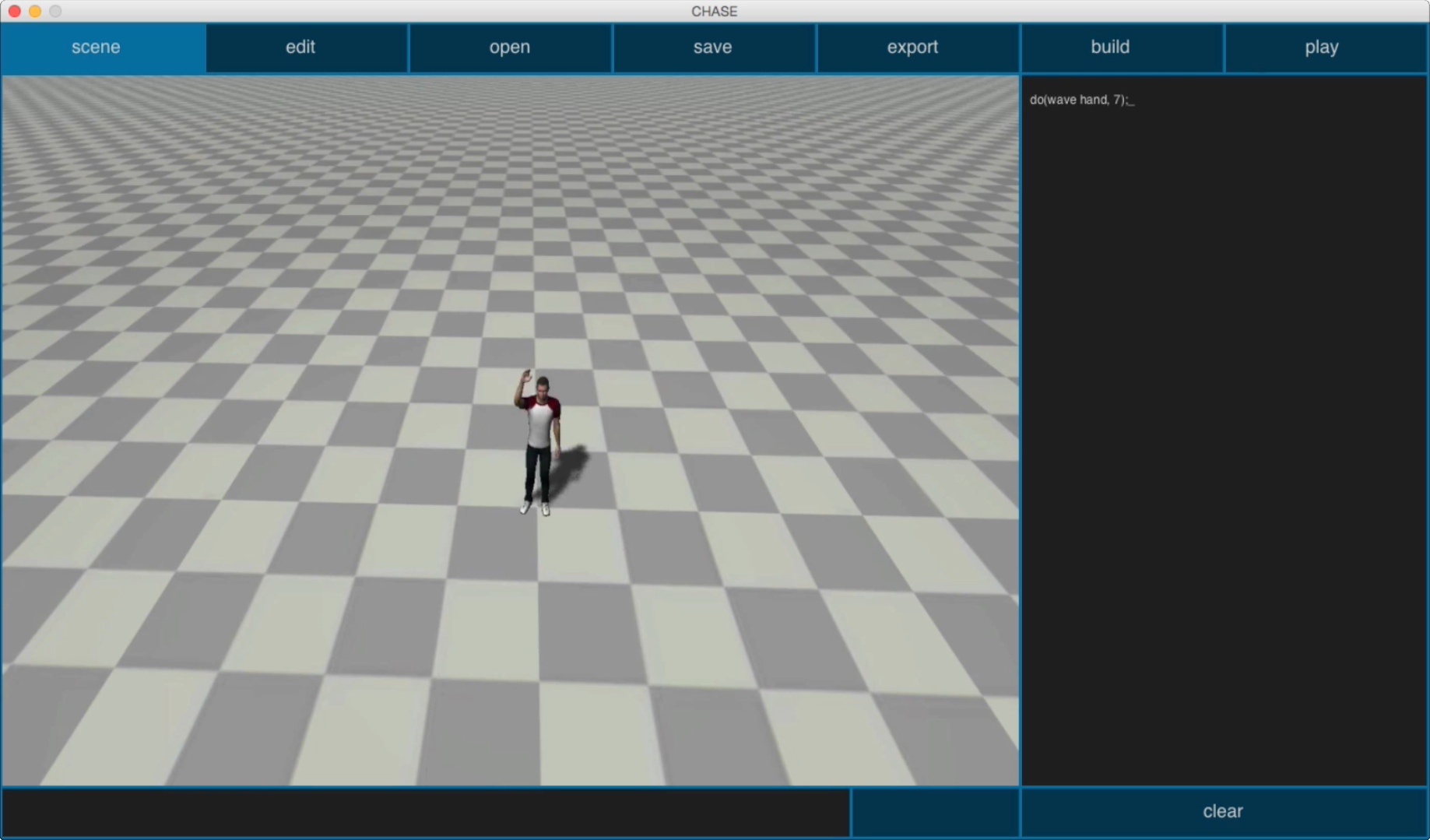}
\caption{The interface of CHASE.}
\label{fig1}
\end{figure*}

In order to implement CHASE the following parts were developed. Firstly, identifying the basic actions that a character should be able to perform and also generating the basic scripting commands. Secondly, a number of parameters that should allow the user not only to synthesize the required motion of a character, but also to gain a higher level of control of each action of the character were defined. By using a reach number of motions that a character can perform, as well as by associating these actions with specified keywords, a motion dataset is created. The input commands are handled by a number of developed background algorithms, which are responsible for retrieving the desired motions and synthesizing the requested actions of the character. During the application's runtime, CHASE synthesizes the requested motion of the character and displays the final animated sequence.

The remainder of this paper is organized as follows. Section 2 covers related work in character animation by presenting previous solutions for animating virtual characters that are based on interactive or automatic techniques. Previously developed scripting environments for the animation of virtual characters are also presented and discussed. A system overview of CHASE is presented in Section 3. The script commands, possible parameters, and additional functionalities that have been developed for CHASE are presented in Section 4. The results, which indicate the potential use of a scripting environment by users who are inexperienced in programming, are presented in Section 5. Finally, conclusions are drawn and potential future work is discussed in Section 6.

\section{1.	Related Work}
\label{sec2}
This section presents work that is related to the solution presented. Specifically, the following paragraphs present methodologies that use different input devices or easily specified constraints for animating virtual characters, systems that provides to a user the ability to synthesize task-based or scenario-related animated sequences, and previously proposed scripting environments for character animation. Finally, the advantages provided by CHASE comparing by previous solutions are presented.

Interactive character control can be classified according to the input device that is used for the character animation process \cite{ref8}. In general, the character controller can be a standard input device, such as a keyboard and a joystick \cite{ref9}. Alternatively, it can be more specialized, such as text input \cite{ref39}\cite{ref40}\cite{ref41}, prosodic features of speech \cite{ref10}, drag and drop systems where the motion sequences are placed into a time-line \cite{ref30}, sketch-based interfaces \cite{ref11} or the body of a user \cite{ref12}\cite{ref42}, while the motion is captured by motion capture technologies. Each of the previously mentioned methodologies has advantages and disadvantages. The choice of the most appropriate input device depends on the actual control of the character's motion that the user requires. 

A variety of methodologies for the animation of a virtual character based on easily specified constraints have also been examined. These solutions are based on motion graphs \cite{ref6}\cite{ref46} literature such as \cite{ref13}, simple footprints \cite{ref14}\cite{ref45} that a character should follow, on space-time constraints as proposed in \cite{ref15}, or statistical models \cite{ref16}\cite{ref49}\cite{ref50} that are responsible for retrieving and synthesizing a character's motion. However, even if easily specified constraints enable a user to animate a character, different frameworks that permit either the interactive or automatic animation of a character have been developed. In \cite{ref17}, which is a task-based character animation system, by using a number of screen buttons, the user is able to animate a character and make it interact with objects that are located within the virtual environment. Other methods, such as \cite{ref18}\cite{ref19}\cite{ref20}, which can be characterized as scenario-based character animation systems, provide automatic synthesizing of a character's motion based on AI techniques.

In the past, researchers developed scripting languages and systems in the field of embodied conversational agents. The XSAMPL3D \cite{ref25}, AniLan \cite{ref27}, AnimalScript \cite{ref28}, SMIL-Agent \cite{ref29} and many others enable a user to script a character's actions based only on predefined command. Among the best-known markup languages for scripting the animation of virtual characters are the Multi-Modal Presentation Markup Language \cite{ref23}, the Character Markup Language \cite{ref24}, the Multimodal Utterance Representation Markup Language \cite{ref25}, the Avatar Markup Language \cite{ref26}, the Rich Representation Language \cite{ref27}, the Behavior Markup Language \cite{ref37} and the Player Markup Language \cite{ref38}, which developed for controlling the behavior of virtual characters.

Various solutions that are similar to the presented methodology were proposed previously for the animation of virtual characters based on scripting commands. StoryBoard \cite{ref21} provides the ability to integrate a scripting language into an interactive character animation framework. Improv \cite{ref22}, another framework with which to create real-time behavior-based animated actors, enables a user to script the specific action of a character based on simple behavior commands. STEP \cite{ref26} framework provides a user the ability to script such actions as gestures and postures. This methodology, which is based on the formal semantics of dynamic logic, provides a solid semantic foundation that enriches the number of actions that a character can perform.

The majority of previously developed scripting environments and markup languages provide only specific actions that a character can perform. An additional limitation is the inability of such systems to enhance a character's synthesized motion. Therefore, a user always receives a lower level of control of the synthesized motion of a character. Moreover, in cases in which a user must generate an animated sequence where many characters will take part, a great deal of effort will be required due to the difficulty of scripting multiple actions for multiple characters. This is especially true for users who wish to generate a sequence with animated characters, but are inexperienced in programming.

These difficulties are overcome in the presented scripting environment. Firstly, instead of enabling a user to script an animated character based on XML-related formats, a simplified scripting environment with its associated scripted language, which is based only on three commands, is introduced. Secondly, since a character should be able to perform concurrent actions, a simple extension of the basic command handles this. Therefore, the user achieves a higher level of control of a character's action. Moreover, in cases where a user must animate more than one character simultaneously, one can specify the character that should perform the requested action by adding an additional method to the existing command for a character. Finally, in cases where a user must generate an animated character in a multitask scenario, by simply specifying the row in which the task should appear, the system will synthesize the tasks requested automatically.

We assume that the described unique functionalities that are implemented in CHASE will enable a user to synthesize compelling animated sequences in which a variety of virtual characters are involved. Hence, in view of the simplicity of the developed commands, in conjunction with the associated parameters, the proposed methodology is quite powerful in comparison to the previous solution. In addition, the easy-to-use and easy-to-remember commands make the presented scripting environment effective, especially for users who are inexperienced in programming.

\section{System Overview}
\label{sec3}
This section briefly describes the proposed system. Specifically, a variety of background algorithms are responsible for recognizing the input commands and synthesizing the motion of a character. The developed background algorithms communicate with the animation system, which is responsible for generating a character's motion, as well as with a path-finding methodology to retrieve the path that the character should follow when a locomotion sequence is required. Finally, CHASE synthesizes and displays the requested motion sequence. Fig. 2 represents the procedure.

\begin{figure}[!t]
\centering
\includegraphics[width=\columnwidth]{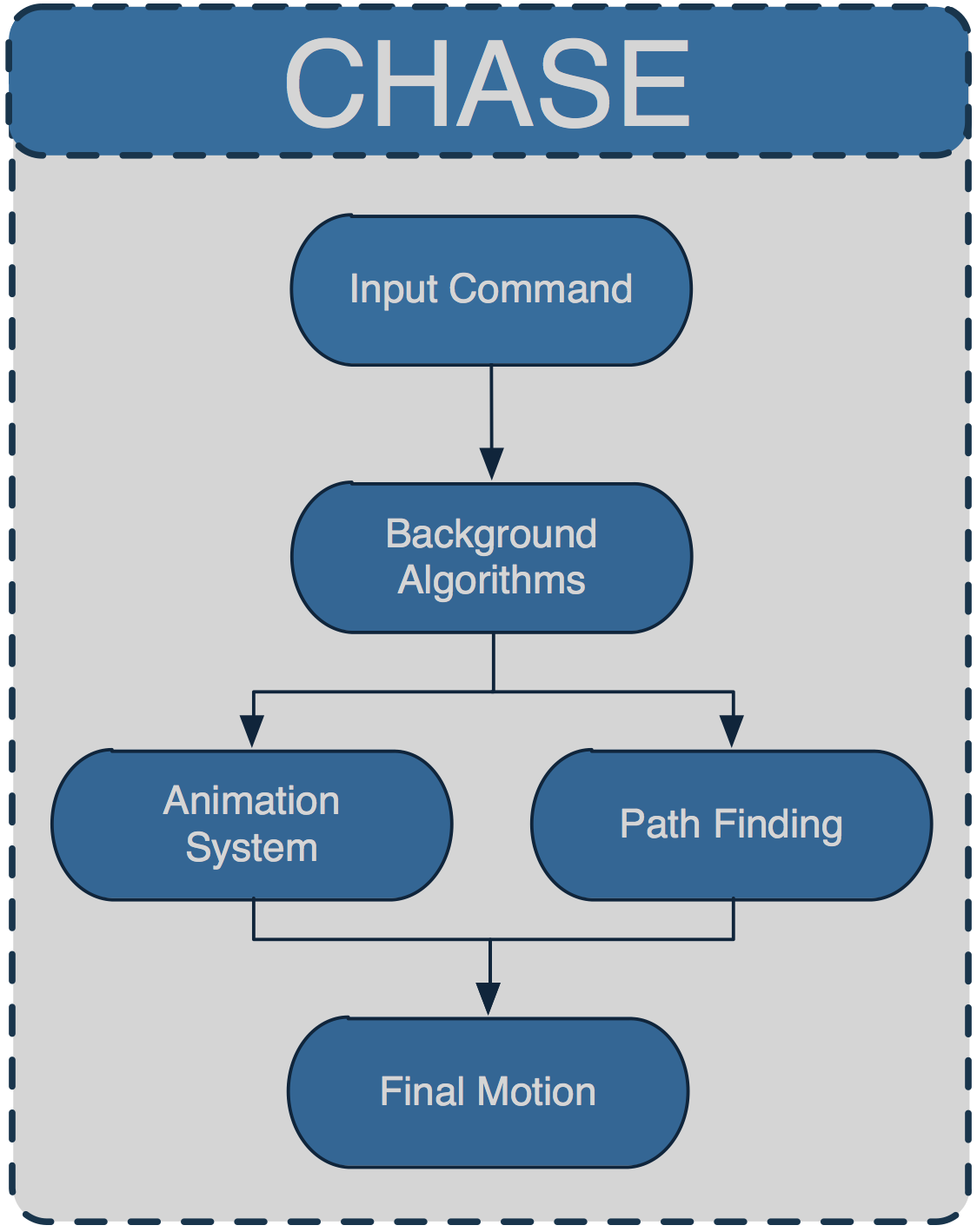}
\caption{The architecture of CHASE.}
\label{fig1}
\end{figure}

\subsection{System Overview}
\label{sec31}
The interface of CHASE (see Fig. 1) is characterized by its simplicity. In its current implementation, it consists of a scene panel that displays the resulting animations, an edit mode panel to edit the input objects, a progress bar that shows the progress of the displayed animation, a scripting box, and a few buttons for use in building, playing and clearing the written scripts. Finally, buttons that save the scripted code and export the generated animated sequences are also provided.
A downloadable version of the presented system, documentation specifying all of its capabilities, and examples of scenes can be found on the CHASE webpage.

\subsection{Third-Party Implementations}
\label{sec32}
A number of techniques and libraries are used to construct CHASE. CHASE uses the Recast/Detour library \cite{ref31} in conjunction with the MEESPM \cite{ref47}\cite{ref48} for the path finding process and collision avoidance with the environment. Concurrent actions are generated based on a simple layering methodology similarly to the one proposed in \cite{ref30}. Finally, a similar to \cite{ref32} full-body inverse kinematics solver was implemented to handle the postures of a character, while interacting with objects located within the virtual environment.

\section{Scripting Character Animation}
\label{sec4}
Developing scripting commands for animating a virtual character can be characterized as a complex process since a virtual character should be able to perform a variety of actions. In this section, the identifications of the basic scripting commands that are necessary to enable the virtual character to navigate and interact within a virtual environment are presented. Moreover, by introducing additional methods called by the main scripts, the system generates concurrent actions of a character, as well as animates multiple characters simultaneously. Finally, an additional functionality of CHASE for scripting multitask animated sequences for the generation of scenario-related animated characters is presented.

\subsection{Identifying Scripting Commands}
\label{sec41}
The application that is presented has been developed for users who are inexperienced in programming. Thus, simple, easily memorized, scripting commands are necessary. To generate the required scripting commands, one must begin by identifying the possible actions or type of actions that a character should perform. Generally, a character should be able to perform simple actions such as waving its hand, tasks related to locomotion such as moving to a target position and interaction tasks such as grasping with its hand an object that is located in the three-dimensional environment. It is apparent that these are the three basic types of actions that a virtual character should be able to perform. Based on this general description, three basic scripting commands were developed: the \texttt{do(parameters)}, the \texttt{goTo(parameters)} and the \texttt{interactWith(parameters)}. 

The \texttt{do(parameters)} command provides a character with the ability to perform a single action. The \texttt{goTo(parameters)} forces a character to move within the given virtual environment. The final command is responsible for making the virtual character capable of interacting with a variety of tasks. Hence, the third command, the \texttt{interactWith(parameters)}, is responsible for providing the ability to control a variety of the character's actions.

For these commands, the \texttt{parameters} within the parentheses indicate the possible parameters that each of the scripting commands could receive (see Section 4.2). Due to the various parameters that each command receives, a user is provided with the means to develop both abstract and specified action of a character. For example, with the \texttt{goTo(parameters)} command, it is possible not only to generate the required locomotion of a character, but also to enable a user to gain better control of the synthesized motion of a character, since the user can specify how the locomotion of a character should be generated. The following section presents the basic parameters that each command receives.

\subsection{Command Parameters}
\label{sec42}

A task assigned to a character can be performed in a variety of different ways. For example, a sequence of locomotion to a target position can be performed by walking, running, etc. motions. Hence, in cases where a user needs a higher level of control of the synthesized motions of a character, parameters that enhance these actual actions generated by the previously mentioned scripting commands should be defined. 

The first command that implemented the \texttt{do(parameters)} command, enables a user  to script simple actions of a character. This command has a single mandatory parameter, which indicates the action that the character should perform. However, optional parameters to specify the body part or the duration of the task can also be used. Specifically, the user can request a single action by calling \texttt{do(action)}, as well as specify the target where the action should be performed, the duration of the action and the body part that should perform the requested action. This command initially permitted a character to perform the requested action without the need to perform a locomotion sequence (i.e., to wave its hand while staying in its position). However, the \texttt{do(parameters)} command can also be used to permit the character to perform locomotion tasks, since one can request that a character perform a walking motion. Based on these parameters that can be inserted into the \texttt{do(parameters)} command, a user has the means not only to generate the requested action, but also to generate an action that should fulfill user-specified constraints.

The \texttt{goTo(parameters)} command enables the character to perform locomotion tasks. The user identifies a mandatory parameter, which is the target position that the character should reach. However, the user is also able to use an additional optional parameter that specifies the motion style that will animate the character. Therefore, a character's locomotion to a target position can be scripted either by (i) inserting the target position such as \texttt{goTo(target)} when a simple walking motion of the character is desired or (ii) inserting \texttt{goTo(target, motion style)} when both target position and motion style are specified.

The final command that is implemented in CHASE, the \texttt{interactWith(parameters)}, can be characterized as more complex than the two previously mentioned commands. The reason is that there are numerous possible interactions between a character and an object. If a character is asked to interact with an object, various actions can be generated. Even if possible to associate actions with specific body parts of a character in a pre-processing stage, there are also possible variations of the required actions. These variations may be related to the character's body or to the duration of the display of the action. For example, scripting a character to kick a ball may also require specifying the foot that should perform this action. Moreover, asking a character to knock a door may also require specifying the duration in the knocking. For that reason, four different parameters have been defined. The first two parameters (\texttt{object name} and \texttt{interaction module}) are mandatory. They indicate the object that the character should interact with, and the interaction module that should be generated. However, depending on the user's requirements for generating a specific action, two more optional parameters could also be inserted. The first one (\texttt{body part}) enables the user to choose which of the character's body parts should perform the requested action. In the current implementation, the user is permitted to choose the hand or foot that will perform the action. The second parameter (\texttt{duration}) enables the user to choose the time (in seconds) required for the requested action.

Based on the possible parameters that each command could receive, the following should be noted. Firstly, while the user did not specify any optional parameter for a scripted command, the system generates the required action taking into account a predefined set of parameters that are associated with each action of the character. For example, if a user requests that a character kick a ball, the system will display only a single kick by the character. The reason is that a ball kicking action is defined as to be performed only once to avoid synthesizing meaningless and repeated motions. Secondly, it should be noted that each optional parameter is independent. This means that the user is not required to specify all of the optional parameters provided by each command. Therefore, the user may control specific components of the requested action. A simple example of this capability of the commands illustrates this. While using the \texttt{do(parameters)} command, the user may request that only either the \texttt{body part} or the \texttt{duration} parameter, or both of these, be filled. In any case, the system's decision in generating the requested motion is not influenced by other factors since it is capable of recognizing the correct form of the scripted command in all of the aforementioned cases.

The three commands that are examined in this paper in conjunction with the associated parameters that can be used to animate a virtual character are summarized in Table \ref{tab1}. In addition, a small set of possible keywords that the user could employ in order to animate virtual characters is presented. It is assumed that an additional control parameter for the synthesized motion could be quite beneficial, since it enables the user not only to animate a character, but also to force the system to synthesize the user's actual wish. Complete documentation of all possible actions that can be synthesized by the character can be found in the CHASE webpage (\emph{ url omitted for review purposes}).

\begin{table*}[!t]
\centering
\includegraphics[width=0.985\textwidth]{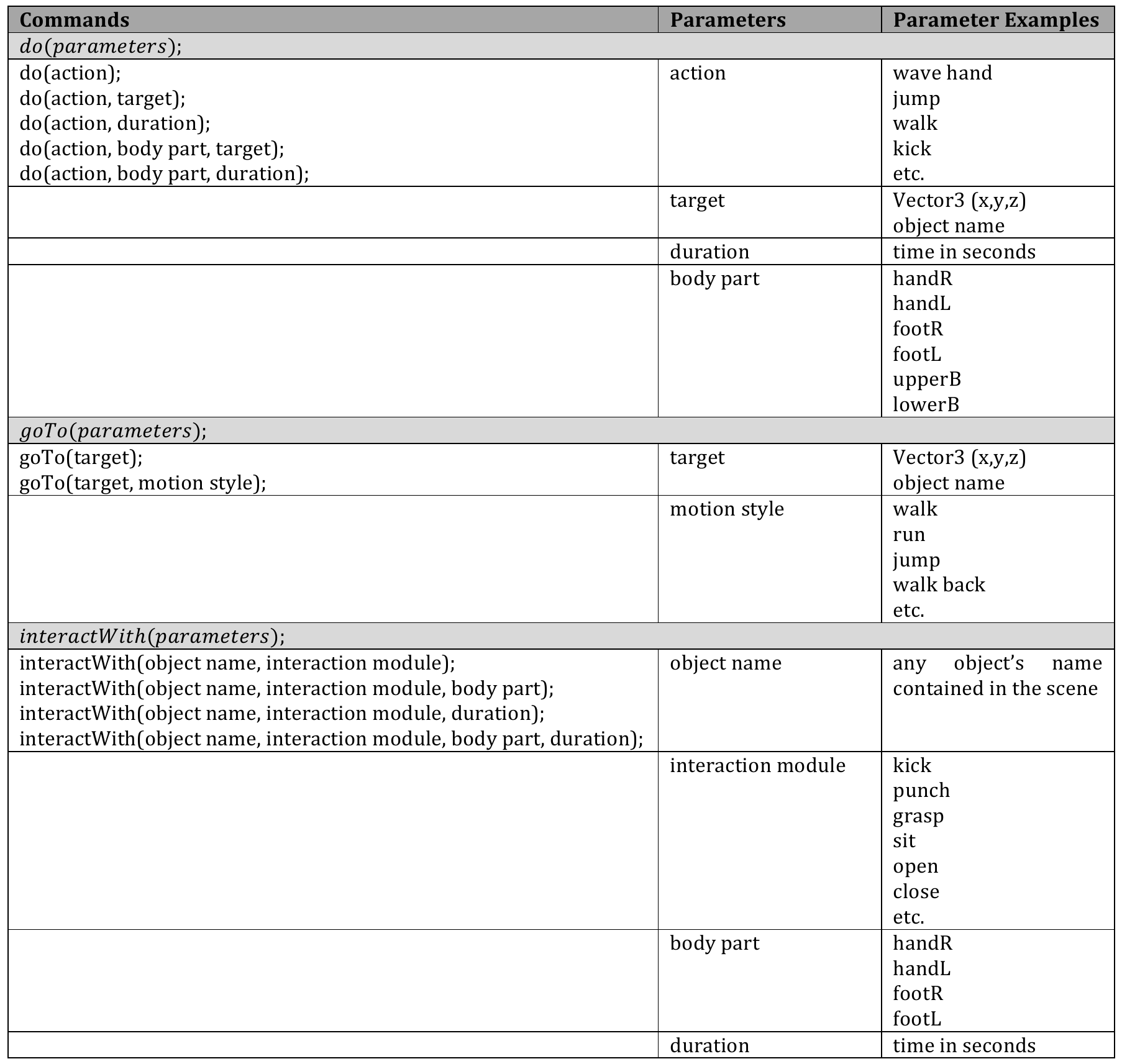}
\caption{\label{tab1} Commands and associated parameters that can be used in CHASE to request an action by an animated virtual character.}
\end{table*}

\subsection{Scripting Concurrent Actions}
\label{sec43}
Virtual characters, such as humans, should be able to perform more than one action simultaneously. This section presents the scripting process for concurrent actions that a character can perform. The concurrent action functionality is based upon the ability to specify the body part that should perform the action in conjunction with the base action that has been requested. The concurrent action lies between the \texttt{do(parameters)} and either the \texttt{goTo(parameters)} or the \texttt{interactWith} \texttt{(parameters)} commands. Specifically, to have a character perform concurrent actions, the \texttt{do(parameters)} command is attached  to either the \texttt{goTo(parameters)} or the \texttt{interactWith} \texttt{(parameters)}. A simple example follows. To cause a character to perform a motion, such as waving its hand while walking to a target position, the system permits the user to script the desired walking motion of a character and to request the additional motion that the system should generate. Hence, the previous example can be requested simply by scripting \texttt{goTo(target, walk).do(wave hand, handR)}. Thus, by permitting the user to generate additional actions of a character, while another action is in progress can, be quite beneficial when more complex animated sequences are required. Therefore, this additional functionality provides a higher level of control over a requested action of a virtual character.

\subsection{Scripting Multiple Characters}
\label{sec44}
In animated sequences it is quite common for more than one character to participate in a single scenario. Hence, by extending the three scripting commands, CHASE also enables a user to script more than one character simultaneously. This is achieved by attaching an additional command to one of the three basic commands, called \texttt{characterName(parameter)}. This command specifies the character that should perform an action, permitting the user to control multiple characters, in cases where more than one character participates in the animation process. A simple example of forcing a specific character to perform an action follows. Consider a character named \texttt{Rudy} who is required to walk to \texttt{target}. This procedure could be called by simply scripting \texttt{goTo(target).characterName(Rudy)}.

\subsection{Scripting Multiple Tasks}
\label{sec45}

In scenario-related sequences that involve virtual characters, the latter should be able to perform a variety of tasks one after the other. Thus, this paper presents a method to script multiple tasks, such as enabling a user to synthesize long animated sequences. Generally, the tasks that a character can perform are characterized by their linearity. Specifically, a task begins while a previous task is completed, and the procedure continues until there are no other tasks for a character to perform.

\begin{figure*}[htb]
  \centering
    \includegraphics[width=0.32\textwidth]{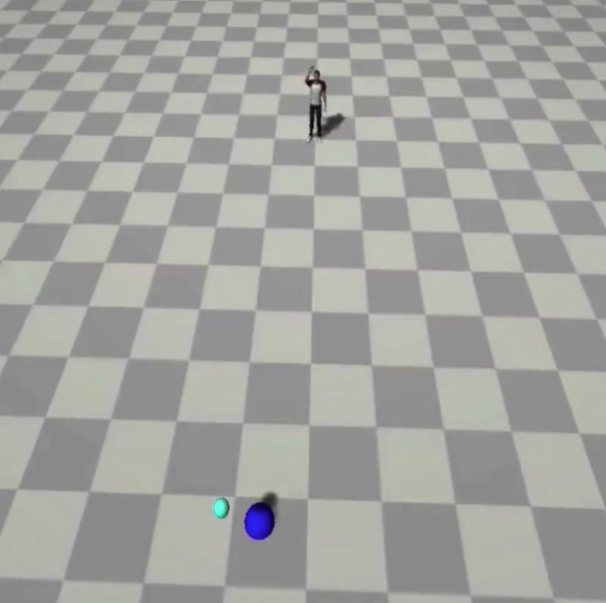}
    \includegraphics[width=0.32\textwidth]{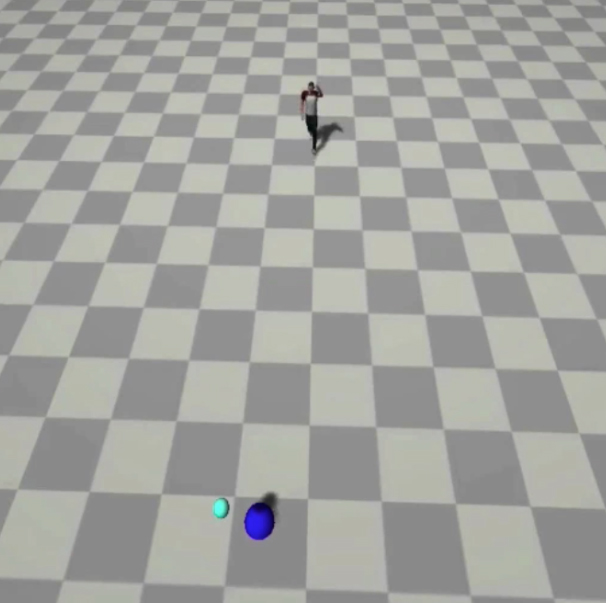}
    \includegraphics[width=0.32\textwidth]{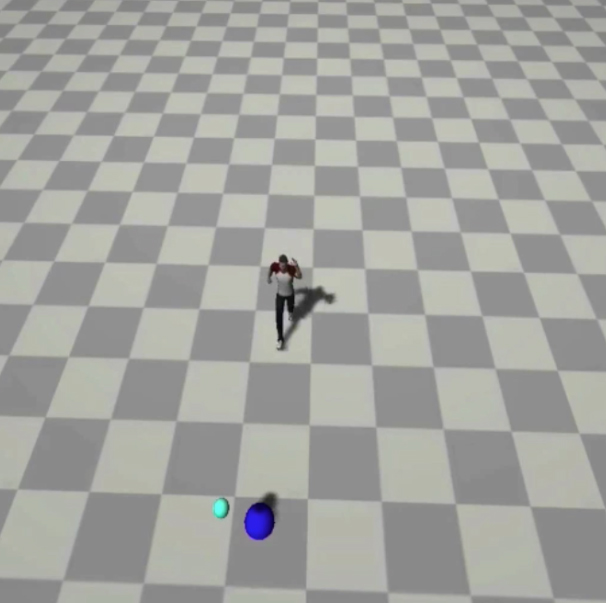}
    \includegraphics[width=0.32\textwidth]{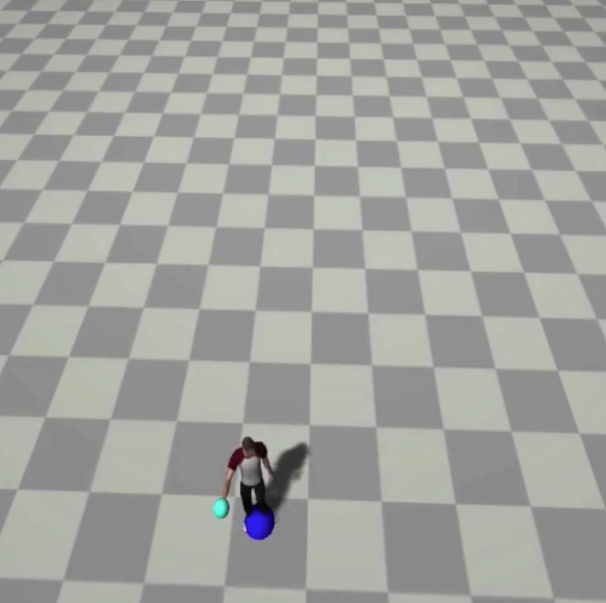}    
    \includegraphics[width=0.32\textwidth]{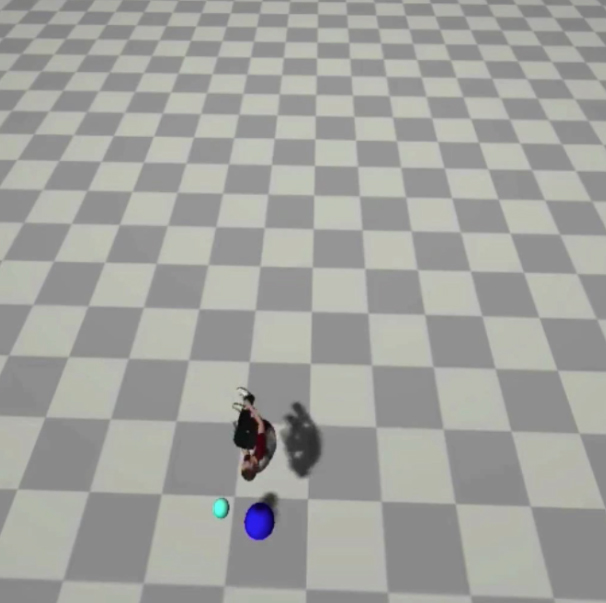}    
    \includegraphics[width=0.32\textwidth]{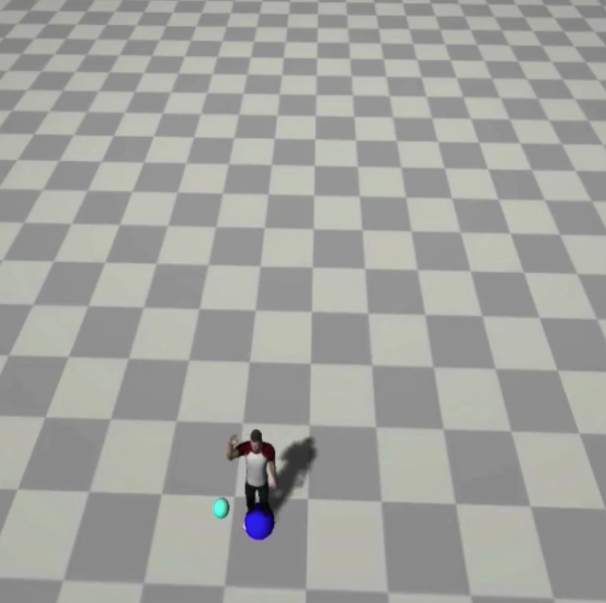}    
	\caption{A multitask scenario generated by using Algorithm \ref{algo1}.}
	\label{fig3}
\end{figure*}

Based on the foregoing, a multitask scenario in a general form can be represented as components of an array that has a dimensionality equal to $N \times 1$, where $N$ denotes the total number of tasks that a character should perform. By assigning each of the actions an array called \texttt{task[index]},  a user can generate long animated sequences. This is achieved by allowing the user to assign singe tasks at each \texttt{index} value of the \texttt{task} array. A simple example of a multitask scenario appears in Figure \ref{fig3}, as well as in the accompanying video. Its scripting implementation is represented in Algorithm \ref{algo1}.

\begin{algorithm}[t]
 \KwData{Input commands of a user}
 \KwResult{The result animated sequence}
task\cite{ref1} = do(wave hand, handR, 3)\;
task\cite{ref2} = goTo(ball, walk).do(wave hand, handL)\;
task\cite{ref3} = interactWith(ball, punch, handR)\;
task\cite{ref4} = do(jump)\;
task\cite{ref5} = do(wave hand, handR, 2)\;
\caption{A simple example for generating a multitask scenario.}
\label{algo1}
\end{algorithm}

\begin{algorithm}[t]
 \KwData{Input commands of a user}
 \KwResult{The result animated sequence}
task\cite{ref1} = goTo(ball, walk).characterName(characterA)\;
task\cite{ref2} = goTo(ball, walk).characterName(characterB)\;
\caption{By placing the actions of two different characters at different \texttt{index} values of the \texttt{task} array, the system generates each character action one after the other.}
\label{algo2}
\end{algorithm}

\begin{algorithm*}[htb]
 \KwData{Input commands of a user}
 \KwResult{The result animated sequence}
tasks\cite{ref1}\cite{ref1} = goTo(target, walk).characterName(characterA)\;
tasks\cite{ref1}\cite{ref2} = goTo(target, run). characterName(characterB)\;
tasks\cite{ref2}\cite{ref1} = interactWith(characterB, punch, handR).characterName(characterA)\;
\caption{A multitask scenario in which there are two characters. In this scenario, \texttt{characterA} moves to its \texttt{target} position while walking, and \texttt{characterB} moves to its \texttt{target} position while running. Finally, \texttt{characterA} punches \texttt{characterB} with his right hand.}
\label{algo3}
\end{algorithm*}

It is quite common in multitask scenarios to involve multiple characters. Two different approaches can be used in CHASE to script more than one character simultaneously in a multitask scenario. The first approach animates each character one after the other. This means that the action required of a character B is generated after the action of a character A has been completed. The reason is that each task of the characters taking part in the multitask scenario have been assigned a different \texttt{index} value of the \texttt{task} array. A simple example of generating the actions of two different characters appears in Algorithm \ref{algo2}. However, a user should be able to animate virtual characters simultaneously in multitask scenarios. This is achieved in CHASE by using a two dimensional array named \texttt{tasks[index][index]}. In this array the first \texttt{index} value represents the row in which each action in generated, whereas the second \texttt{index} value represents the number of the character. It should be noted that each character should be represented by the same \texttt{index} value while developing a multitask scenario. Hence, the background algorithms that are implemented recognize and generate the requested tasks as separate entries. This enables the user to animate a number of characters simultaneously. A simple example in which there are two characters in a multitask scenario appears in Algorithm \ref{algo3}. It should be noted that a multitask scenario where multiple characters evolve in a general form can be represented as an array that has a dimensionality equal to $M \times N$, where $M$ denotes the total number of characters evolving in the multitask scenario and $N$ denotes the total number of tasks that a character should perform.

\section{Evaluation and Results}
\label{sec5}
This section covers the results obtained while evaluating CHASE. To understand the efficiency of using CHASE, we designed a study to evaluate the ability of individual students to animate virtual characters that perform specified tasks. The participants were asked to use CHASE to develop a specified scenario. Two different teams of participants took place in the evaluation process. The first team consisted of eight participants who were first or second year undergraduate students in science or engineering departments. These students had a basic knowledge of computer programming (had attended only an introductory course in computer programming). However, none of them had previous experience in, or any particular knowledge of, computer graphics and animation. The second team consisted of eight participants who were in their first or second year of study in arts. None of the students of this team had any prior knowledge of programming. All participants can be characterized a computer literate.

\subsection{Experimental Methodology}
\label{sec51}
Each participant sat down in a quiet room in front of a desktop computer in which the CHASE application had been installed. First, the experimenter demonstrated the functionalities of CHASE for 10 minutes. During that time, each participant had the opportunity to note freely. After the demonstration, each participant was given the opportunity to discuss the system and to ask questions. An additional five minutes were provided for experimentation with CHASE. However, no further explanation of the system was given.

After this introductory procedure, the experimenter explained to the participants that they would use CHASE to generate a scenario with only one character involved. Specifically, it was asked to generate the scenario presented in Algorithm 1. The scenario was provided in a printed form in which the characters' actions were described sequentially in writing. The actions that each character was required to perform and the position that was to be reached were written in bold letters in the printed document. The basic commands and parameters that the users were required to employ were supplied on a separate sheet. The time (in seconds) that was required to complete the given scenario was recorded. However, each participant was permitted a maximum of 15 minutes to complete the assigned scenario. After a participant completed the assigned scenario, he or she evaluated the difficulty of generating an animated sequence using the commands and functionalities of CHASE by assigning a score of 1 (difficult) to 7 (easy-to-use). Finally, since CHASE was developed to enable inexperienced programmers to visualize and generate their own stories that involved virtual characters, the participants were asked to evaluate CHASE by assigning a score of 1 (would not like to use it again) to 7 (I would strongly like to use it again).

\subsection{Results}
\label{sec52}
We recorded separately for each team the time that participants took to complete the given scenario. A pairwise t-test revealed that science and engineering students completed the given scenario more quickly than the students from the arts department. Specifically, the mean time for the science and engineering students was estimated at $m=316.9$ seconds with a standard deviation of $\sigma=69.2$. The arts department students had a mean time of $m=657.2$ seconds with a standard deviation of $\sigma=86.7$. Based on these results it can be stated that students more familiar with programming are able to finish the given task faster. However, by considering the achievement of the arts department students of completion of the given tasks, it could be stated that also inexperienced in programming users can efficiently use such a scripting environment.

For evaluation purposes, we also recorded the students' satisfaction in interacting with CHASE by scoring the difficulty of using the developed commands in CHASE. Users were invited to evaluate the intuitiveness and difficulty of each command on a 7-point Likert scale (1 being difficult, 7 being easy). A pairwise t-test found that it was easier for participants from the science and engineering departments to use the scripting commands than it was for the students from the arts department. Specifically, for the arts department students $t(8) = 3.09$, $p=0.018$, with the science and engineering department students $m=4.6$ on the difficulty scale and $\sigma=1.3$, while in the arts department students $m=3.8$ with $\sigma=1.7$. However, a pairwise t-test did not reveal any statistical difference between the two intuitiveness scales ($p>0.05$).

The final results indicate the potential willingness of each participant to use CHASE in the future. Surprisingly, all participants responded that they would like to use CHASE in the future. Specifically, $m=6.75$ with $\sigma=0.46$. Therefore, it is shown that all of the participants would like to visualize their ideas and also develop their own stories that involve virtual characters by using CHASE.

\section{Conclusions and Future Work}
\label{sec6}
In this paper, a novel scripting environment, called CHASE, for use in animating virtual characters was presented. CHASE enables a user to request a variety of actions that a character can perform by simply using three commands. Each command, which receives a variety of parameters, is associated with specific actions that the character is able to perform. Moreover, the commands communicate with a variety of background algorithms that are responsible for generating the actions requested of the character. In addition to the scripting commands, by introducing three additional functionalities, the user is able to script concurrent actions of a character, multiple characters at the same time, and multi-task scenarios in order to generate scenario-related sequences that involve animated characters. To demonstrate the efficiency and simplicity of use of CHASE, an evaluation process was conducted. Two teams of students took part in the study. The first team consisted of students who had minimal previous programming experience, whereas the second team consisted of students who had no previous programming experience. This evaluation process has shown how easy it is to use such a scripting environment for the animation of virtual characters, as well as the potential willingness of users to employ CHASE in order to visualize their ideas and produce their own stories that involve animated virtual characters.

In its current version, CHASE provides a variety of functionalities. However, there are additional functionalities that we would like to implement in the near future. Specifically, we would like to implement additional commands in conjunction with the associate parameters to allow the system to provide complex interaction between multiple characters. In addition, we would like to expand the concurrent actions functionality by allowing the user to specify more than one body part that performs additional actions simultaneously. Moreover, in the current version of CHASE, the camera that is used to capture the virtual content is placed into the virtual environment in a specific position. Hence, in the future we would like to provide functionalities to allow the camera to be positioned by scripts, as well as to animate in response to the users' requests. We assume that the additional functionalities mentioned will be quite beneficial in generating a wide range of actions and interaction, as well as in representing the generated sequences more properly. As a result, users would be able to generate compelling sequences that involve animated virtual characters.

\bibliographystyle{ieeetran}
\bibliography{IEEEabrv}

\begin{thebibliography}{10}
\providecommand{\url}[1]{#1}
\csname url@samestyle\endcsname
\providecommand{\newblock}{\relax}
\providecommand{\bibinfo}[2]{#2}
\providecommand{\BIBentrySTDinterwordspacing}{\spaceskip=0pt\relax}
\providecommand{\BIBentryALTinterwordstretchfactor}{4}
\providecommand{\BIBentryALTinterwordspacing}{\spaceskip=\fontdimen2\font plus
\BIBentryALTinterwordstretchfactor\fontdimen3\font minus
  \fontdimen4\font\relax}
\providecommand{\BIBforeignlanguage}[2]{{%
\expandafter\ifx\csname l@#1\endcsname\relax
\typeout{** WARNING: IEEEtran.bst: No hyphenation pattern has been}%
\typeout{** loaded for the language `#1'. Using the pattern for}%
\typeout{** the default language instead.}%
\else
\language=\csname l@#1\endcsname
\fi
#2}}
\providecommand{\BIBdecl}{\relax}
\BIBdecl

\bibitem{ref1}
M.~Gleicher, ``Retargetting motion to new characters,'' in \emph{25th Annual
  Conference on Computer Graphics and Interactive Techniques}.\hskip 1em plus
  0.5em minus 0.4em\relax New York, USA: ACM Press, July 1998, pp. 33--42.

\bibitem{ref2}
A.~Witkin and Z.~Popovic, ``Motion warping,'' in \emph{22nd Annual Conference
  on Computer Graphics and Interactive Techniques}.\hskip 1em plus 0.5em minus
  0.4em\relax New York, USA: ACM Press, September 1995, pp. 105--108.

\bibitem{ref3}
L.~Kovar and M.~Gleicher, ``Flexible automatic motion blending with
  registration curves,'' in \emph{ACM SIGGRAPH/Eurographics Symposium on
  Computer Animation}.\hskip 1em plus 0.5em minus 0.4em\relax UK: Eurographics
  Association, July 2003, pp. 214--224.

\bibitem{ref4}
S.~I. Park, H.~J. Shin, and S.~Y. Shin, ``On-line locomotion generation based
  on motion blending,'' in \emph{ACM SIGGRAPH/Eurographics Symposium on
  Computer Animation}.\hskip 1em plus 0.5em minus 0.4em\relax UK: Eurographics
  Association, July 2002, pp. 105--111.

\bibitem{ref5}
B.~Van~Basten and A.~Egges, ``Motion transplantation techniques: A survey,''
  \emph{Computer Graphics and Applications}, vol.~32, no.~3, pp. 16--23, 2012.

\bibitem{ref43}
C.~Mousas and P.~Newbury, ``Real-time motion synthesis for multiple
  goal-directed tasks using motion layers,'' in \emph{Virtual Reality
  Interaction and Physical Simulation}, 2012, pp. 79--85.

\bibitem{ref44}
C.~Mousas, P.~Newbury, and C.-N. Anagnostopoulos, ``Splicing of concurrent
  upper-body motion spaces with locomotion,'' \emph{Procedia Computer Science},
  vol.~25, pp. 248--359, 2012.

\bibitem{ref6}
L.~Kovar, M.~Gleicher, and F.~Pighin, ``Motion graphs,'' \emph{ACM Transactions
  on Graphics}, vol.~21, no.~3, pp. 473--482, 2002.

\bibitem{ref7}
T.~Mukai and S.~Kuriyama, ``Geostatistical motion interpolation,''
  \emph{Transactions on Graphics}, vol.~24, no.~3, pp. 1062--1070, July 2005.

\bibitem{ref8}
N.~Sarris and M.~G. Strintzis, \emph{3D Modeling and Animation: Synthesis and
  Analysis Techniques for the Human Body}.\hskip 1em plus 0.5em minus
  0.4em\relax Hershey, Pennsylvania: IGI Global, July 2003.

\bibitem{ref9}
J.~McCann and N.~Pollard, ``Responsive characters from motion fragments,''
  \emph{ACM Transactions on Graphics}, vol.~26, no.~3, p. Article No. 6, August
  2007.

\bibitem{ref39}
M.~Oshita, ``Generating animation from natural language texts and semantic
  analysis for motion search and scheduling,'' \emph{The Visual Computer},
  vol.~26, no.~5, pp. 339--352, 2010.

\bibitem{ref40}
C.~Mousas and C.-N. Anagnostopoulos, \emph{Character Animation Scripting
  Environment}, ser. Encyclopedia of Computer Graphics and Games.\hskip 1em
  plus 0.5em minus 0.4em\relax Springer, 2015.

\bibitem{ref41}
C.~Mousas and C.-N. Anagnostopoulos, ``Chase: Character animation scripting
  environment,'' in \emph{VRCAI}, 2015, pp. 55--62.

\bibitem{ref10}
S.~Levine, C.~Theobalt, and V.~Koltun, ``Real-time prosody-driven synthesis of
  body language,'' \emph{ACM Transactions on Graphics}, vol.~28, no.~5, p.
  Article No. 28, 2009.

\bibitem{ref30}
M.~Oshita, ``Smart motion synthesis,'' \emph{Computer Graphics Forum}, vol.~27,
  no.~7, pp. 1909--1918, October 2008.

\bibitem{ref11}
J.~Davis, M.~Agrawala, E.~Chuang, Z.~Popovi{\'c}, and D.~Salesin, ``A sketching
  interface for articulated figure animation,'' in \emph{ACM
  SIGGRAPH/Eurographics Symposium on Computer Animation}.\hskip 1em plus 0.5em
  minus 0.4em\relax UK: Eurographics Association, July 2003, pp. 320--328.

\bibitem{ref12}
J.~Chai and J.~K. Hodgins, ``Performance animation from low-dimensional control
  signals,'' \emph{ACM Transactions on Graphics}, vol.~24, no.~3, pp. 686--696,
  2005.

\bibitem{ref42}
C.~Ouzounis, C.~Mousas, C.-N. Anagnostopoulos, and P.~Nebury, ``Using
  personalized finger gestures for navigating virtual characters,'' in
  \emph{Virtual Reality Interaction and Physical Simulation}, 2015, pp. 5--14.

\bibitem{ref46}
C.~Mousas and P.~Newbury, ``Real-time motion editing for reaching tasks using
  multiple internal graphs,'' in \emph{IEEE Computer Games Conference}, 2012,
  pp. 51--55.

\bibitem{ref13}
A.~Safonova and J.~K. Hodgins, ``Construction and optimal search of
  interpolated motion graphs,'' \emph{ACM Transactions on Graphics}, vol.~26,
  no.~3, p. Article No. 106, August 2007.

\bibitem{ref14}
M.~Van De~Panne, ``From footprints to animation,'' \emph{Computer Graphics
  Forum}, vol.~16, no.~4, pp. 211--223, October 1997.

\bibitem{ref45}
C.~Mousas, P.~Newbury, and C.-N. Anagnostopoulos, ``Footprint-driven locomotion
  composition,'' \emph{International Journal of Computer Graphics and
  Animation}, vol.~4, no.~4, pp. 27--42, 2014.

\bibitem{ref15}
M.~F. Cohen, ``Interactive spacetime control for animation,'' \emph{ACM
  SIGGRAPH Computer Graphics}, vol.~26, no.~2, pp. 293--302, July 1992.

\bibitem{ref16}
J.~Min and J.~Chai, ``Motion graphs++: A compact generative model for semantic
  motion analysis and synthesis,'' \emph{ACM Transactions on Graphics},
  vol.~31, no.~6, p. Article No. 153, 2012.

\bibitem{ref49}
C.~Mousas, P.~Newbury, and C.-N. Anagnostopoulos, ``Data-driven motion
  reconstruction using local regression models,'' in \emph{Artificial
  Intelligence Applications and Innovations}, 2014, pp. 364--374.

\bibitem{ref50}
C.~Mousas, P.~Newbury, and C.-N. Anagnostopoulos, ``Evaluating the covariance
  matrix constraints for data-driven statistical human motion reconstruction,''
  in \emph{Spring Conference on Computer Graphics}, 2014, pp. 99--108.

\bibitem{ref17}
A.~W. Feng, Y.~Xu, and A.~Shapiro, ``An example-based motion synthesis
  technique for locomotion and object manipulation,'' in \emph{ACM SIGGRAPH
  Symposium on Interactive 3D Graphics and Games}.\hskip 1em plus 0.5em minus
  0.4em\relax New York, USA: ACM Press, March 2012, pp. 95--102.

\bibitem{ref18}
M.~Thiebaux, S.~Marsella, A.~N. Marshall, and M.~Kallmann, ``Smartbody:
  Behavior realization for embodied conversational agents,'' in
  \emph{International Joint Conference on Autonomous Agents and Multiagent
  Systems}, vol.~1, International Foundation for Autonomous Agents and
  Multiagent Systems.\hskip 1em plus 0.5em minus 0.4em\relax New York, USA: ACM
  Press, May 2008, pp. 151--158.

\bibitem{ref19}
M.~Kapadia, S.~Singh, G.~Reinman, and P.~Faloutsos, ``A behavior-authoring
  framework for multiactor simulations,'' \emph{Computer Graphics and
  Applications}, vol.~31, no.~6, pp. 45--55, 2011.

\bibitem{ref20}
A.~Shoulson, N.~Marshak, M.~Kapadia, and N.~I. Badler, ``Adapt: The agent
  development and prototyping testbed,'' in \emph{ACM SIGGRAPH Symposium on
  Interactive 3D Graphics and Games}.\hskip 1em plus 0.5em minus 0.4em\relax
  New York, USA: ACM Press, March 2013, pp. 9--18.

\bibitem{ref25}
A.~Vitzthum, H.~B. Amor, G.~Heumer, and B.~Jung, ``Xsampl3d: An action
  description language for the animation of virtual characters,'' \emph{Journal
  of Virtual Reality and Broadcasting}, vol.~9, p. Article No. 1, 2012.

\bibitem{ref27}
A.~Formella and P.~P. Kiefer, ``Anilan - an animation language,'' in
  \emph{Computer Animation}.\hskip 1em plus 0.5em minus 0.4em\relax New York,
  USA: IEEE, June 1996, pp. 184--189.

\bibitem{ref28}
G.~R{\"o}{\ss}ling and B.~Freisleben, ``Animalscript: An extensible scripting
  language for algorithm animation,'' \emph{ACM SIGCSE Bulletin}, vol.~33,
  no.~1, pp. 70--74, February 2001.

\bibitem{ref29}
K.~Balci, E.~Not, M.~Zancanaro, and F.~Pianesi, ``Xface open source project and
  smil-agent scripting language for creating and animating embodied
  conversational agents,'' in \emph{International Conference on
  Multimedia}.\hskip 1em plus 0.5em minus 0.4em\relax New York, USA: ACM Press,
  September 2007, pp. 1013--1016.

\bibitem{ref23}
H.~Prendinger, S.~Descamps, and M.~Ishizuka, ``Mpml: A markup language for
  controlling the behavior of life-like characters,'' \emph{Journal of Visual
  Languages \& Computing}, vol.~15, no.~2, pp. 183--203, 2004.

\bibitem{ref24}
Y.~Arafa and A.~Mamdani, ``Scripting embodied agents behaviour with cml:
  Character markup language,'' in \emph{International Conference on Intelligent
  User Interfaces}.\hskip 1em plus 0.5em minus 0.4em\relax New York, USA: ACM
  Press, January 2003, pp. 313--316.

\bibitem{ref26}
Z.~Huang, A.~Eli{\"e}ns, and C.~Visser, ``Step: A scripting language for
  embodied agents,'' in \emph{Workshop of Lifelike Animated Agents}.\hskip 1em
  plus 0.5em minus 0.4em\relax Berlin, Germany: Springer-Verlag, August 2002,
  pp. 87--109.

\bibitem{ref37}
H.~Vilhjalmsson, N.~Cantelmo, J.~Cassell, N.~E. Chafai, M.~Kipp, S.~Kopp,
  M.~Mancini, S.~Marsella, A.~N. Marshall, C.~Pelachaud, Z.~Ruttkay, K.~R.
  Thorisson, H.~v. Welbergen, and R.~J. V.~D. Werf, ``The behavior markup
  language: Recent developments and challenges,'' in \emph{Intelligent virtual
  agents}.\hskip 1em plus 0.5em minus 0.4em\relax Springer Berlin-Heidelberg,
  January 2007, pp. 99--111.

\bibitem{ref38}
Y.~A. Jung, ``Animating and rendering virtual humans: Extending x3d for real
  time rendering and animation of virtual characters,'' in \emph{International
  Conference on Computer Graphics Theory and Applications}.\hskip 1em plus
  0.5em minus 0.4em\relax UK: SCITEPRESS, 2008, pp. 387--394.

\bibitem{ref21}
M.~Gervautz and D.~Schmalstieg, ``Integrating a scripting language into an
  interactive animation system,'' in \emph{Computer Animation}.\hskip 1em plus
  0.5em minus 0.4em\relax New York, USA: IEEE Press, May 1994, pp. 156--166.

\bibitem{ref22}
K.~Perlin and A.~Goldberg, ``Improv: A system for scripting interactive actors
  in virtual worlds,'' in \emph{23rd Annual Conference on Computer Graphics and
  Interactive Techniques}.\hskip 1em plus 0.5em minus 0.4em\relax New York,
  USA: ACM Press, August 1996, pp. 205--216.

\bibitem{ref31}
M.~Mononen, ``Recast/detour navigation library,''
  \url{https://github.com/memononen/recastnavigation}, accessed 29/11/2014.

\bibitem{ref47}
C.~Mousas, P.~Newbury, and C.-N. Anagnostopoulos, ``Rethinking shortest path:
  An energy expenditure approach,'' in \emph{Virtual Reality Interaction and
  Physical Simulation}, 2013, pp. 35--39.

\bibitem{ref48}
C.~Mousas, P.~Newbury, and C.-N. Anagnostopoulos, ``The minimum energy
  expenditure shortest path method,'' \emph{Journal of Graphics Tools},
  vol.~17, no. 1-2, pp. 31--44, 2013.

\bibitem{ref32}
P.~Lang, ``Root-motion,'' \url{http://www.root-motion.com/}, accessed
  29/11/2014.

\end{thebibliography}

\end{document}